\documentclass[onecolumn]{emulateapj}
\usepackage{amsmath, amsthm, amssymb}

\newcommand{\imin}{i_{\rm min}}
\newcommand{\mt}{M_{\rm T}}
\newcommand{\mo}{M_{\rm obs}}
\newcommand{\mm}{m_{\rm min}}
\newcommand{\mx}{m_{\rm max}}

\slugcomment{ApJ, accepted}

\shorttitle{Effect of viewing angle}
\shortauthors{Lopez \& Jenkins}

\begin{document}

\title{ The effects of viewing angle on
  the  mass distribution of exoplanets}

\author{S. Lopez \& J.S. Jenkins}
\affil{Departamento de Astronom\'ia, Universidad de Chile,
    Casilla 36-D, Santiago, Chile}

\email{slopez@das.uchile.cl}

\begin{abstract}

We present a mathematical method to statistically decouple the effects of
unknown inclination angles on the mass distribution of exoplanets that have
been discovered using radial-velocity techniques. The method is based on the
distribution of the product of two random variables. Thus, if one assumes a
true mass distribution, the method makes it possible to recover the observed
distribution.  We compare our prediction with available radial-velocity
data. Assuming the true mass function is described by a power-law, the minimum
mass function that we recover proves a good fit to the observed distribution
at both mass ends. In particular, it provides an alternative explanation for
the observed low-mass decline, usually explained as sample incompleteness.  In
addition, the peak observed near the the low-mass end arises naturally in the
predicted distribution as a consequence of imposing a low-mass cutoff in the
true-distribution.  If the low-mass bins below 0.02M$_{\rm{J}}$ are complete,
then the mass distribution in this regime is heavily affected by the small
fraction of lowly inclined interlopers that are actually more massive
companions. Finally, we also present evidence that the exoplanet 
mass distribution changes form towards low-mass, implying that a single 
power law may not adequately describe the sample population.

\end{abstract}


\section{Introduction}

Samples of exoplanetary systems are increasing rapidly thanks to new ground
and space-based dedicated surveys, thus enabling investigation of their
statistical properties. One of these properties is the planetary mass
distribution, a key aspect needed to understand the origin of exoplanets and
its relation to the initial mass function. Currently, radial-velocity (RV)
detections (e.g. \citealp{mayor83}; \citealp{butler96}; \citealp{jones10})
have provided the largest sample of unconstrained systems. However, the RV
technique does not provide masses directly because the line-of-sight
inclination angles, $i$, cannot be measured unless complementary observations
are carried out, for instance transit photometry (\citealp{henry00}) or
astrometry (\citealp{benedict06}). Thus, all masses measured with this
technique are indeed 'minimum' planet masses, $\mo=\mt\sin{i}$, where $\mt$,
the 'true' planet mass, is not known a priori.

Understanding the \emph{true} mass distribution rather than the minimum mass
distribution will allow modelers to compare their mass distributions against
a function that is free from one of the largest sources of uncertainty (see
\citealp{mordasini09}; \citealp{ida05}).  Also, the $\sin i$ degeneracy that
plagues RV signals means we can never be fully sure that any
individual signature is planetary in nature from the Doppler data alone.  This
has consequences for a number of aspects of planetary, brown dwarf, and
low-mass star studies that deal with inferences drawn from a RV
dominated mass distribution.  A prime example of this would be the proper
location in mass of the planet-brown dwarf boundary (see
\citealp{sahlmann11}), which allows one to clarify the status of an object as
either a planet or a brown dwarf (e.g. \citealp{jenkins09}), and will help us
to better understand the formation mechanisms of both classes of objects.

It is thought that the $\sin{i}$ correction is of order unity and
would preserve the power-law shape of the observed mass distribution (e.g.,
\citealp{jorissen01}; \citealp{tabachnik02}; \citealp{hubbard07}; Morton \&
Johnson 2011). However, no proof has been provided {to substantiate}
this. Methods to recover the $\mt$ distribution from the observed
  minimum-mass data have been proposed based on: (1) numerically solving an
  Abel-type integral equation that relates observed and true mass
  distributions (\citealp{jorissen01}); (2) analytically finding the
  distribution that maximizes the likelihood of having a given set of minimum
  masses (\citealp{zucker01}); (3) comparing cumulative distributions of
  projected and de-projected data with non-parametric statistical tools
  (\citealp{brown11}); and (4) using the physics of multi-planet systems to
  resolve the $\sin{i}$ correction (\citealp{batygin11}). With the exception
  of (4), which is a theoretical prediction, these methods are non-parametric,
  i.e., they do not assume an a priori model of the data. However, they also suffer
  from some drawbacks like the need to smooth the data (1 and 3), or the
  complexity of introducing observational limits (2).

In this paper we present an alternative method to statistically decouple the
$\sin i$ dependence in observed exoplanet mass functions.  The method is based
on the expected distribution of the product of two continuous and independent
random variables.  Its parametric nature requires an assumption on the shape
of the underlying ($\mt$) distribution; but, on the other hand, it offers the
possibility of introducing observational constraints in a straightforward
fashion. The mathematical problem, applied to planet mass distributions, is
stated in \S~\ref{problem}, and its solution presented in
\S~\ref{solution}. In \S~\ref{examples} the method is implemented on two
example distributions and in \S~\ref{predictions} we make a comparison with
observational data.  In \S~\ref{discussion} we look at the significance of the
shape of the true mass distribution and how this may affect current models of
planet formation and evolution. Finally, in \S~\ref{summary} we summarize the
results.

\section{The problem and our concept}
\label{problem}

The problem can be summarized as follows: given an analytical model of the
distribution of true planetary masses, is it possible to obtain the
distribution of minimum-masses analytically by assuming a random distribution
of inclination angles? The answer to this question is yes, and relies on
computing the probability density function (PDF) of the product of two random
variables. The problem of finding the PDF of the product of two random
variables was first solved by Rohatgi (1976) but its implementation relied on
knowledge of the joint PDF of the two variables.  In this paper we apply the
approach followed by Glen et al. (2004), which uses the {\it individual} PDFs 
of the two random variables.

To do so, let $X$, $Y$, and $Z$ be random variables, such that 

\begin{equation}
X=\mt
\end{equation}
\begin{equation}
Y=\sin{i}
\end{equation}
\begin{equation}
Z=XY~.
\end{equation}

The above equations state that $X$ takes the value of any exoplanetary mass,
$\mt$, $Y$ takes the value of any correction by inclination (viewing) angle
$i$, $\sin{i}$, and $Z$ takes the value of any possible product
$\mt\sin{i}$. Let also $f_X$, $f_Y$, and $f_Z$ be their respective probability
density functions (PDF). We wish to obtain $f_Z$, a prediction for the
observed distribution of 'minimum' masses, given $f_X$ (an assumption for the
true mass distribution) and $f_Y$ (which can be calculated analytically).

\section{Solution}
\label{solution}

\begin{figure}
\center
\includegraphics[angle=-90,scale=0.6,clip]{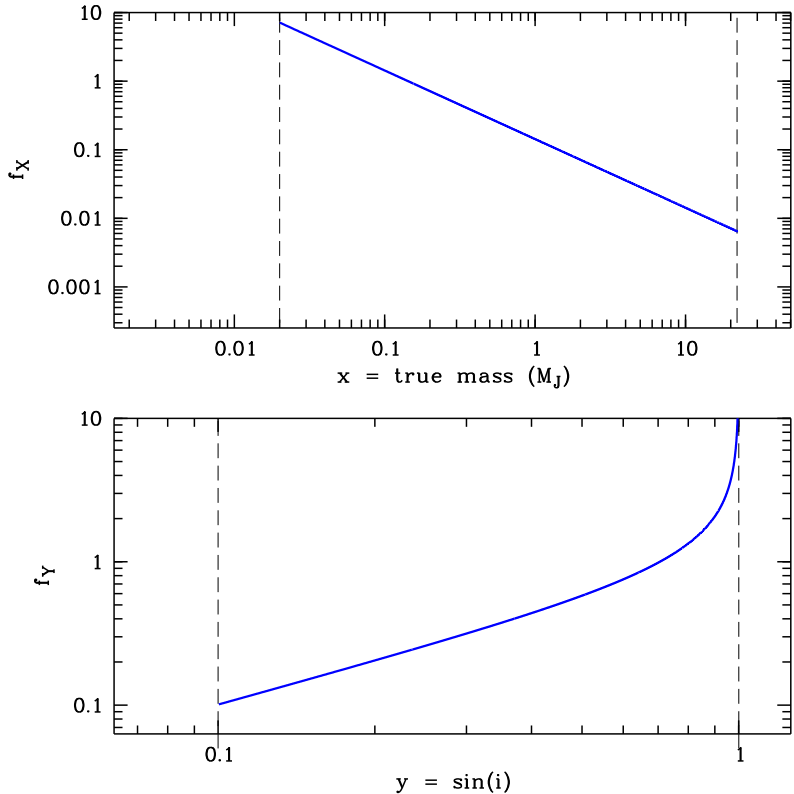}
\caption{Top panel: PDF of $X$ describing true masses, with $\alpha=0$$,
  \mm=0.02$ and $\mx=22$ M$_{\rm J}$. Bottom panel: PDF of $Y$ describing
  $\sin i$ corrections, with $i$ values drawn from a uniform distribution in
  the range [5.7,90.0] degrees.
\label{fig_fXfY}
} 
\end{figure}

\subsection{PDF of $X$}
\label{pdf_x}

Let $f_X$ represent the distribution of planet true masses. For simplicity, we
assume $f_X$ is well described by a power-law. We define a power-law of index
$-(1+\alpha)$, such that it is normalized over the mass interval $[\mm,\mx]$,
defined to be the minimum and maximum true masses.  Note that a minimum
non-zero mass avoids a diverging integral, which is a condition for $f_X$ to
be normalized. Thus, $f_X$ becomes

\begin{equation}
f_X(x)~dx=\begin{cases}
A_X x^{-\alpha - 1}~dx, & 0<\mm<x<\mx\\
0,&\text{otherwise,}
\end{cases}
\label{eqX}
\end{equation}

where $x$ represents true masses,
$A_X\equiv\alpha/(\mm^{-\alpha}-\mx^{-\alpha})$ is a normalization constant
and $\alpha>-1$. If $\alpha=0$, then $A_X=(\ln{(\mx/\mm)})^{-1}$. It is
important to note that $X$ can be treated as a {\it continuous} variable and
that the present analysis does not require that masses be normalized (i.e.,
$x$ can be in any units).  An example of this PDF, representing the true mass
distribution, is shown in the top panel of Fig.~\ref{fig_fXfY}.

\subsection{PDF of $Y$}
\label{pdf_y}

Consider randomly distributed inclination angles, $i$. The {\it observed}
inclination angles can be shown to be distributed like $\sin{i}$, over the
interval $[0,\frac{\pi}{2}]$.  This is straightforward to see in spherical
coordinates (e.g., Ho \& Turner 2011) and implies that higher inclination
angles (edge-on) are more probable than lower ones (pole-on).

To get $f_Y$ for a $\sin{i}$-distribution of angles, let
$y=\sin{i}$. Then $i=\arcsin{y}$, with $i \in [0,\frac{\pi}{2}]$. The
corresponding cumulative distribution (CDF) is given by $1-\cos{i}$. Thus, the
differential of this CDF, $d(1-\cos{[i]})=d(1-\cos{[\arcsin{y}]})$ gives
$f_Y$:

\begin{equation}
f_Y(y)~dy=\frac{y}{\sqrt{1-y^2}}~dy, \hspace{1cm}
0<y<1.
\end{equation}

The above equation considers all possible angles. However, it may be
useful to define a minimum inclination angle, $i_{\rm min}$, to account for
possible selection effects when comparing with real data. In this case
$f_Y$ becomes

\begin{equation}
f_Y(y)~dy=A_Y\frac{y}{\sqrt{1-y^2}}~dy, \hspace{1cm}
\sin{i_{\rm min}}<y<1.
\label{eqY}
\end{equation}

where $A_Y\equiv 1/\sqrt{1-\sin{(i_{\rm min})}^2}$ is a  normalization
constant. An example of this PDF, representing  the 
$\sin i$ distribution, is shown in the bottom panel of Fig.~\ref{fig_fXfY}

\subsection{PDF of $Z$}
\label{sectionfz}

Having defined $f_X$ and $f_Y$, the PDF of random variables $X$ and $Y$, we
can now calculate $f_Z$, the PDF of the product $XY$. Following Glen et
al. (2004)'s solution for the PDF of the product of two continuous and
independent random variables, and also considering the above boundaries,
$f_Z(z)$ can be expressed as

\begin{equation}
f_Z(z)= \begin{cases}
\int\limits_{\mm}^{z/\sin{i_{\rm min}}} f_Y(\frac{z}{u}) f_X(u) \frac{1}{u}du, &
 \mm\sin{i_{\rm min}}<z<\mm,\\
\int\limits_{z}^{z/\sin{i_{\rm min}}} f_Y(\frac{z}{u}) f_X(u) \frac{1}{u}du, &  
\mm<z<\mx\sin{i_{\rm min}},\\
\int\limits_{z}^{\mx}{f_Y(\frac{z}{u}) f_X(u) \frac{1}{u}du}, &  
\mx\sin{i_{\rm min}}<z<\mx,
\end{cases}
\label{eqZ}
\end{equation}

provided that $\mm<\mx\sin{i_{\rm min}}$. This condition is the equivalent of
setting a lower limit on $i$, such that pole-on orbits, producing very large
corrections, are excluded from the observed sample. It also determines three
'validity regions'.

Replacing Eq.~\ref{eqZ} with Equations~\ref{eqX}
and~\ref{eqY}, after some algebra and getting rid of the integration
limits for a moment, Eq.~\ref{eqZ} reads:

\begin{equation} 
f_Z(z)=
A_X A_Y z\int\frac{u^{-\alpha-2}}{\sqrt{u^2-z^2}}du~. 
\label{eqZ2}
\end{equation}

For any value of $\alpha$, the improper integral in Eq.~\ref{eqZ2} has a
primitive in terms of $_2F_1$, the first hypergeometric function (Abramowitz
\& Stegun 1964). From Eq.~\ref{eqZ}, note also that one important
property of $f_Z$ is that it vanishes at the boundaries $z=\mm\sin{\imin}$ and
$z=\mx$, meaning that the observed mass distribution must have a peak. This
has important consequences when interpreting distributions of real data (see
\S~\ref{discussion}).

In conclusion, the problem stated in \S~\ref{problem} is formally solved for
the distributions described in \S~\ref{pdf_x} and \S~\ref{pdf_y}. Indeed,
Eq.~\ref{eqZ} is valid for any true mass distribution (i.e., not only for
a power-law) but in general the
evaluation of $f_Z$ will require numerical integration.

\section{Examples}
\label{examples}

\begin{figure}
\center
\includegraphics[angle=-90,scale=0.6,clip]{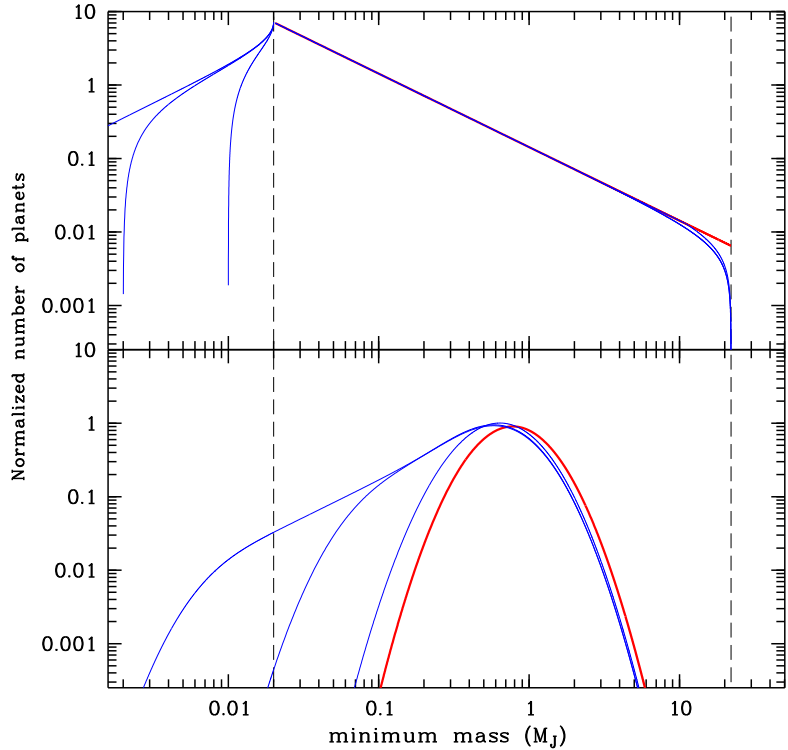}
\caption{
Top panel: power-law distribution of true masses ($f_X$; red curve) with
$\mm=0.02$ M$_{J}$ and $\mx=22$ M$_{J}$, and corresponding predictions of minimum-mass
distributions ($f_Z$; blue curves) for various minimum inclination angles,
$\imin=0.57$, $5.7$, and $30.0$ degrees. Bottom panel: same as above but for a
log-normal distribution of true masses, $f_X(x)=
\frac{1}{x\sigma\sqrt{2\pi}}\exp{-\frac{(\ln{x}-\mu)^2}{2\sigma^2}}$ with $\mu=0$ and $\sigma=0.5$. 
\label{example}
} 
\end{figure}

We now evaluate numerically two examples of true mass distributions: a
power-law and a log-normal distribution.

For the power-law introduced in \S\ref{pdf_x} and used to derive
Eq.~\ref{eqZ2}, the case $\alpha=0$ is easy to evaluate (and will allow us to
perform a quick comparison with RV data in the next Section). Replacing with
$\alpha=0$ and integrating Eq.~\ref{eqZ2}, the predicted minimum-mass
distribution becomes:

\begin{equation} 
f_Z(z)dz=\frac{A_XA_Y}{z}\times
\begin{cases}
\left[\frac{\sqrt{(z/\sin{\imin})^2-z^2}}{z/\sin{\imin}} -
  \frac{\sqrt{\mm^2-z^2}}{\mm} \right]dz, &
 \mm\sin{i_{\rm min}}<z<\mm,\\
\left[\frac{\sqrt{(z/\sin{\imin})^2-z^2}}{z/\sin{\imin}}\right]dz, &  
\mm<z<\mx\sin{i_{\rm min}}, \\
\left[\frac{\sqrt{\mx^2-z^2}}{\mx}\right]dz,  &  
\mx\sin{i_{\rm min}}<z<\mx.\\
\end{cases}
\label{eqZback}
\end{equation} 

The function $f_Z$ represents the predicted distribution of minimum-masses 
if the true mass distribution is proportional to $\mt^{-1}$.

The top panel of Fig.~\ref{example} shows $f_Z$ for various minimum
inclination angles (blue curves; $\imin=0.57$, $5.7$, and $30.0$ degrees). For
reference, the underlying distribution of true masses, $f_X$, is also shown
(red curve), where we have set $\mm=0.02$ M$_{J}$ and $\mx=22$ M$_{J}$. The
effect of the $\sin{i}$ correction is evident at both mass ends of the
predicted distribution, as objects in any mass bin 'migrate' to lower-mass
bins in the minimum mass distribution.
  
However, the shape of the low-mass tail ($z<\mm$) is affected mostly by the
inclusion of large $\sin i$ corrections, i.e., lowly-inclined systems. This is
readily seen from the fact that $f_Z$ converges to $f_X$ in the limit
$\sin{\imin}\approx 1$. Thus, in general, {\it the observed distribution of
a true mass power-law distribution (with boundaries) should show a decrease at
the low-mass end.}

The same effects are observed if a log-normal distribution is used instead of
a power-law.  The bottom panel of Fig.~\ref{example} shows such a distribution
(same color codes for the true and predicted mass distributions and same
$\imin$ values), where Eq.~\ref{eqZ} has been integrated numerically using 

\begin{equation} 
f_X(x)=\frac{1}{x\sigma\sqrt{2\pi}}\exp{-\frac{(\ln{x}-\mu)^2}{2\sigma^2}}
\end{equation} 

with $\mu=0$ and $\sigma=0.5$. 

We conclude by emphasizing that setting a particular value of $\imin$
simulates observational selection effects, meaning that the model mimics
observational samples which have missed objects at angles below that
limit. Fig.~\ref{example} shows that including those lowly-inclined systems,
despite them being less probable, induces dramatic changes in the shape of the
predicted distribution at the low-mass end.

We now proceed to compare our models with RV data.

\section{Comparison with observational data}
\label{predictions}

\begin{figure}
\center
\includegraphics[angle=-90,scale=0.6,clip]{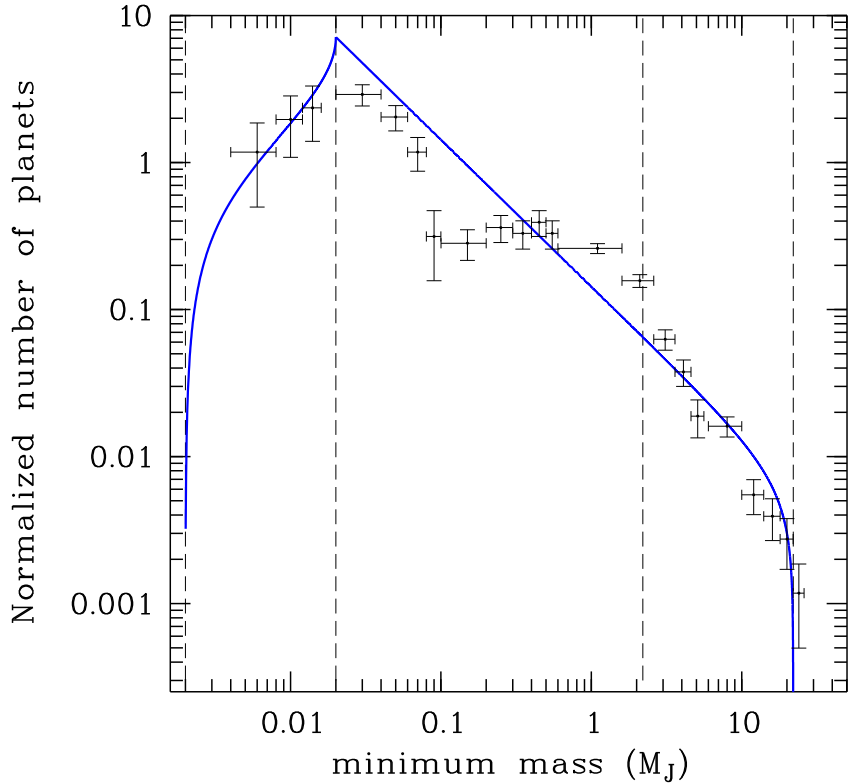}
\caption{
Observed distribution of minimum masses for a sample of 643 RV discovered
exoplanets (data taken from {\tt http://exoplanet.eu} as of October 17
2011). Bin sizes were arbitrarily chosen to have a roughly equal number of
systems. Overlaid is the model of minimum masses given by Eq.~\ref{eqZback}
and obtained from an assumed true mass distribution with the following
parameters: $f(\mt)\propto \mt^{-1}$, $\mm=0.02$ M$_{\rm J}$, $\mx=22$ M$_{\rm
  J}$. The predicted minimum-mass distribution was set to have
$\sin{\imin}=0.1$. The dashed lines mark the validity regions in
Eq.~\ref{eqZback} }
  \label{fig_back}
\end{figure}

Fig.~\ref{fig_back} displays the observed distribution of minimum masses
(Schneider et al. 2011; RV data on 643 planets as of October 17 taken from
{\tt http://exoplanet.eu}). Bin sizes were arbitrarily chosen to have a
roughly equal number of systems. Overlaid is the model prediction of minimum
masses given by Eq.~\ref{eqZback}, i.e., the prediction that results from
assuming a true mass distribution of the form $\propto \mt^{-1}$. We have
chosen the following physical parameters for the true mass distribution:
$\mm=0.02$ M$_{J}$ and $\mx=22$ M$_{J}$, where the lower limit is chosen from
observation and the upper limit marks the planetary-brown dwarf mass
boundary (Sahlmann et al. 2011). To simulate observational selection effects, $\sin{\imin}=0.1$ was
chosen. We emphasize that these parameters are not the result of a fit, but
were adjusted manually to attempt a best match to the data.

It can be seen that the power-law part of the predicted curve
(Fig.~\ref{fig_back}) only poorly fits the data. However, interestingly
enough, both of its extremes do seem to better reproduce the data.

At the low-mass limit ($z<0.02$ M$_{\rm J}$) the model describes mass-bins
which are not 'occupied' in the true mass distribution. As already
  mentioned, the shape of this tail is affected mostly by the inclusion of
  large $\sin i$ corrections, i.e., lowly-inclined systems, which in turn
  produce a decrease at the low-mass end of the observed distribution. Such a
decrease mimics observational effects on the data like those ones induced by
incompleteness (\citealp{udry07}), as we discuss further in next Section.

At the high-mass end, there is no apparent reason for a systematic lack of
observed systems. Here the data shows a decline which is also in good
agreement with our model's prediction. However, the small number of
systems precludes any firm conclusion.

An interesting feature that appears in the data between the 
high and low-mass ends of the distribution is a deficit that is not 
well described by the power law model prediction.  Around 0.1 M$_{\rm J}$ 
the data drop and then quickly turnover to rise again, creating a mass 
distribution paucity in the data.  Given we expect this mass regime to be 
fairly well sampled by current RV surveys, this feature may indicate that more 
than a single power law is needed to describe the full ensemble of planetary 
masses.

We conclude by emphasizing that the above approach provides a direct
comparison with observational data. Assuming the true mass distribution is
at least fairly well described by a power-law, we have shown that our solution for the
minimum-mass distribution provides a good fit to the observations at both mass
ends.

\section{Discussion}
\label{discussion}

We have shown that if the true mass distribution is described by a power-law
with boundaries, there must be a peak in the observed mass distribution of
exoplanets, and therefore this peak has implications for planet formation and
evolution models.  A peak in the planetary mass distribution tells us that
most of the mass of the proto-planetary disk that goes into forming planets,
gets locked up in the formation of the larger gas and ice giants.  This agrees
with the mass distribution of the planets in the solar system.

The widely accepted planetary formation theory is by core accretion and
subsequent planet migration (\citealp{pollack84}; \citealp{lin86}) and this
model can broadly explain the currently observed population of exoplanets.
The peak in the planetary mass distribution needs to be taken into account
when comparing the outcomes from core accretion formation models against the
observed mass distribution of exoplanets, unless the true mass distribution
changes form towards the lowest masses.

One interesting question that arises is, what does the position of the peak in
the mass distribution tell us and what does it mean?  As we have seen in
Fig.~\ref{fig_back}, our true mass distribution model can provide a good fit
to the observed mass distribution of the current population of exoplanets,
particularly the peak and subsequent decline.

The low-mass peak we find that best describes the observed data is located
around 0.02M$_{\rm{J}}$, or $\sim$6.5M$_{\oplus}$.  When we look at systems
that have high inclinations, like an observed mass distribution drawn from 
transiting planets only, we find that for a complete
sample, the observed distribution follows the true distribution with no
low-mass peak.  Therefore, if we assume that the bins are complete below
0.02M$_{\rm{J}}$, then this is the regime where we begin to observe the
effects of systems with low inclination angles, and hence large mass
corrections.

Here we are assuming we have sampled all angles above a certain limit,
  $\imin$, but our model considers the lower likelihood of observing systems
with low inclinations; therefore this tells us something important that even a
small fraction of systems with low inclinations can produce large changes in
the observed mass distribution in the low-mass regime. This result is in
  line with the implications of the analysis by Ho \& Turner (2011). These
  authors demonstrate, using Bayesian analysis, that the {\it posterior}
  distribution of angles is determined by the particular true mass
  distribution and so the latter cannot be simply obtained from the observed
  one. Unlike Ho \& Turner, we deal here with the {\it prior} distribution of
  angles and study its effects on the {\it assumed} shape of the true mass
  distribution; however, both approaches lead to the conclusion that the low
  inclination systems modify the low-mass end of the observed distribution.

Also, since most of the observed distribution above the 0.02M$_{\rm{J}}$
boundary broadly follows the true distribution then small to medium values of
$\sin i$ do not affect the overall mass distribution in a large manner.  We
make it clear that most of the low-mass systems in these bins are genuine
rocky planets, but that the small numbers of low-inclination/high-mass
interlopers cause a dramatic change to the observed mass distribution, again,
assuming that the low-mass bins are complete.

Finally, we do caution that the true mass distribution we are discussing here
applies to planets with small semimajor axes ($\le$4~AU's or so).  The
distribution of mass at ever increasing distances from the central star may
change the shape of the true mass distribution, but further analysis on this
issue is likely to require many more detections like the directly imaged
planets around HR8799 (\citealp{marois08}) or planets discovered by
microlensing techniques (e.g. \citealp{muraki11}).

\rm

\section{Summary and outlook}
\label{summary}

 We have applied the formal solution for the PDF of the product of two
 independent random variables to the observational problem of decoupling the
 $\sin i $ factor from an observed  sample of exoplanet  masses. Our
 approach requires that the true mass distribution is modeled by a
 continuous function that represents the PDF (within given physical or
 observational limits).

We have shown that if the true mass
function is modeled as a power-law, comparison with observed data of 643 RV
planets shows a good match with our method's prediction, specifically at both
mass ends. In particular, the prediction agrees well with the decline observed
toward the low-mass end, thus providing an alternative explanation to the
turnover being the result of observational biases.  

If the low-mass bins below 0.02M$_{\rm{J}}$ are assumed to be complete, then 
we show that the presence of a small number of systems with low inclinations, 
and hence much larger true masses, heavily affects the distribution in this 
regime.  Such effects are not seen above this mass region where the $\sin i$ 
values are more modest and hence corrections are smaller, meaning the true 
distribution is being matched more closely.

We also again note a mass paucity around 0.1 M$_{\rm J}$ in the observed mass
distribution, as the single power law model does not describe this region very
well at all.  In fact, it may be prudent to examine the mass distribution using
more than one function, like a double power law for instance.  This may 
indicate that the mass distribution changes form below around 0.1 M$_{\rm J}$,
a feature we plan to study more in the future.

In summary, we have provided a practical and intuitive method to decouple the
$\sin i$ effect that is inherent to RV samples. The method  offers the
possibility of introducing observational constraints in a straightforward
fashion, in order to compare predictions with current observations.

Current RV surveys are plagued with biases and incompleteness that affect
conclusions drawn from analyzing any Doppler data set.  For instance, the RV
detection method is heavily biased towards the detection of more massive
companions on short period orbits, since they induce a larger reflex motion on
the host star in comparison with less massive and longer period companions.
Therefore, the detection of very low-mass planets is only now being fully
realized and requires large data sets and novel detection/characterization
methods (e.g. \citealp{vogt10}; \citealp{pepe11}; \citealp{anglada-escude12}; 
\citealp{jenkins12}).  Hence, studying the
low-mass end of the mass distribution is one of the corner-stones of exoplanet
research at the present time and will continue to be so in the near future.

\acknowledgments

The authors acknowledge the very helpful discussions with Hugh Jones and Raul
Gouet, as well as the important feedback given by an anonymous referee. SL has
been supported by FONDECYT grant number 1100214. JSJ acknowledges funding by
FONDECYT through grant 3110004 and partial support from the Gemini-CONICYT
Fund and from the Comit\'e Mixto ESO-Gobierno de Chile. Wolfram Mathematica
online integrator was used.

\appendix

\end{document}